\documentclass[5p,12pt]{elsarticle}

\usepackage{graphicx}

\usepackage{amssymb}

\biboptions{sort&compress}

\newcommand{\Tr}{{\rm Tr\,}}

\journal{Physics Letters B}

\begin{document}

\begin{frontmatter}



\title{SU($N_c$) gauge theories at deconfinement}


\author{Biagio Lucini}
\ead{B.Lucini@swansea.ac.uk}
\address{College of Science, Swansea University, Singleton Park,
Swansea SA2 8PP, UK}

\author{Antonio Rago}
\ead{antonio.rago@plymouth.ac.uk}
\address{School of Computing and Mathematics, University of Plymouth,
Plymouth PL4 8AA, UK}

\author{Enrico Rinaldi}
\ead{E.Rinaldi@sms.ed.ac.uk}
\address{SUPA and The Tait Institute, School of Physics and Astronomy, University of Edinburgh,
Edinburgh EH9 3JZ, UK}

\begin{abstract}
The deconfinement transition in SU($N_c$) Yang--Mills is
investigated by Monte Carlo simulations of the gauge theory
discretized on a spacetime lattice. We present new results  for $
4 \le N_c \le 8$ (in particular, for $N_c = 5$ and $N_c = 7$), which are
analysed together with previously published results.
The increased amount of data, the improved statistics and simulations
closer to the continuum limit provide us with better control
over systematic errors. After performing the thermodynamic limit,
numerical results for the ratio of the
critical temperature $T_c$ over the square root of the string tension $\sqrt{\sigma}$
obtained on lattices with temporal extensions $N_t = 5,6,7,8$ are
extrapolated to the continuum limit. The continuum results at fixed
$N_c$ are then extrapolated to $N_c = \infty$. We find
that our data are accurately described by the formula
$T_c/\sqrt{\sigma} = 0.5949(17) + 0.458(18)/N_c^2$. 
Possible systematic errors affecting our calculations are also discussed.
\end{abstract}


\end{frontmatter}


\section{Introduction}
\label{sect:introduction}
In recent years, various lattice studies have been performed for SU($N_c$)
gauge theories in the 't Hooft limit (see
e.g. Refs.~\cite{Teper:2009uf,Narayanan:2009xh} for a review).
As a result, on the one hand the old idea that the bulk of the physics is
shared between SU(3) and the simpler
SU($\infty$)~\cite{Hooft:1973jz,Witten:1979kh,Manohar:1998xv} has been 
confirmed; on the other hand, solid bases have been provided for
gauge-string duality studies aiming at describing QCD-like theories
(see e.g. Ref.~\cite{Aharony:1999ti} for an early review of this field). 

On the lattice, one of the main areas of activity has been the finite-temperature
regime~\cite{Lucini:2002ku,Lucini:2003zr,Lucini:2004yh,DelDebbio:2004rw,deForcrand:2004jt,Bursa:2005yv,Lucini:2005vg,Bringoltz:2005rr,Bringoltz:2005xx,deForcrand:2005rg,Liddle:2005qb,Liddle:2008kk,Kiskis:2009xj,Panero:2009tv,Datta:2009jn,Caselle:2011fy,Caselle:2011mn,Mykkanen:2012ri}. In
particular, it has been shown that the deconfinement 
temperature can be determined with very good
accuracy~\cite{Lucini:2002ku,Lucini:2003zr,Lucini:2004yh,Liddle:2005qb,Liddle:2008kk,Datta:2009jn}.
This suggests to use the deconfinement temperature as the physical scale
in large $N_c$ limit studies of observables at fixed lattice
spacing~\cite{Lucini:2004my,DelDebbio:2007wk,Armoni:2008nq,Lucini:2010nv,Lucini:2010kj},
which are often a useful intermediate step before performing the
continuum extrapolation. The main motivation of this work is to
complement existing results on the deconfinement phase transition
by providing the value of the (pseudo-)critical coupling at various
temporal extensions for the gauge groups SU(5) and SU(7), which have
not been investigated at finite temperature before. Some of these
results have already been used in our study of glueball masses at
large $N_c$ at the critical coupling for a temporal extension $N_t$
of six lattice spacings~\cite{Lucini:2010nv}. There, the inclusion of
the $N_c = 5$ and $N_c = 7$ 
data allowed us to increase the precision of the large $N_c$
extrapolation of the masses. The results of that study suggest that knowing the
critical coupling at several values of $N_t$ also for $N_c=5,7$ can improve large
$N_c$ extrapolations in the continuum limit. 

In addition, the calculations presented in this work provide us with an opportunity to
revisit the extrapolation of the critical temperature to $N_c =
\infty$ in the continuum limit. Besides adding the new results to
existing lattice data, we investigate possible systematic
errors. In particular, the finite-size studies of
Refs.~\cite{Lucini:2003zr,Lucini:2005vg} have been performed 
on $N_t = 5$ lattices and the results have been used to perform the
extrapolation at other values of $N_t$ with fixed spatial size
$N_s$. Although this procedure is well justified in principle, there
is the danger that, since the critical coupling at $N_t = 5$ for $N_c \ge
6$ is close to the bulk phase transition, the obtained value for the
coefficient of the leading correction in $1/N_s^3$ in the
thermodynamic extrapolation is significantly
affected by finite lattice spacing artefacts. Since the determination
of that coefficient is performed at a unique $N_t$, if the proximity
to the bulk phase is a problem, the determination of the critical coupling
in the thermodynamic limit at larger $N_t$, and as a consequence the
continuum limit of the critical temperature, will be affected by a
systematic error. In the same spirit, we have performed calculations
at $N_t = 7$ for $4 \le N_c \le 8$, so that the continuum limit can be
obtained by extrapolating the numerical data for the four values of
$N_t = 5, 6, 7, 8$. This allows us to estimate the influence of the
$N_t = 5$ point on the obtained numerical value at zero lattice
spacing, and hence to check whether the
continuum limit is affected by lattice artefacts related to the bulk
phase transition or to the use of too coarse a lattice
spacing in the extrapolation procedure. Finally, we have performed high statistics 
calculations in SU(8), in most cases consisting of at least one million
measurements, in order for the system at criticality to perform at
least 8 round trips ({\em tunnellings}) between the confined and the
deconfined phase. The same criterion in terms of number of tunnellings
has been used to determine the statistics for all the new simulations
discussed in this Letter. This should remove any bias related to a
possible loss of ergodicity in the critical region.   

The rest of this Letter is organised as follows. In
Section~\ref{sect:phasetransitions} we describe the system under
investigation, define the observables we
study and provide numerical results for an estimator of the coupling at which the
deconfinement phase transition takes place at fixed
$N_c$ , $N_t$ and $N_s$. Section~\ref{sect:thermodynamic} deals with the thermodynamic
limit of the critical couplings. Section~\ref{sect:continuum} is devoted to the continuum extrapolation of the
critical temperature in units of the string tension at fixed
$N_c$. The large $N_c$ limit of the latter quantity is discussed
in Section~\ref{sect:tclargen}. Finally, in Section~\ref{sect:conclusions} we summarise
our results and briefly discuss possible future directions. 
\section{The phase transition}
\label{sect:phasetransitions}
Our calculation follows the method exposed
e.g. in Refs.~\cite{Lucini:2002ku,Lucini:2003zr}, which we will briefly
summarise below. We consider a SU($N_c$) gauge theory described by the Wilson action
\begin{equation}
  \label{eq:action-lattice}
  S = \beta \sum_{i,\mu > \nu} \left( 1 - \frac{1}{N_c} \mbox{Re} \ \Tr
    \left( U_{\mu \nu}(i) \right) \right)
  \ ,
\end{equation}
where $U_{\mu \nu}(i)$ is the path-ordered product of the links
$U_{\mu}(i) \in$ SU($N_c)$  around the lattice plaquette identified by the point $i$
and the directions $\mu$ and $\nu$. $\beta$ is defined as $\beta = 2N_c/g_0^2$,
with $g_0$ the bare gauge coupling. The finite-temperature regime is
realised by considering the system on a lattice of volume $L_s^3
\times L_t$, where $L_s = a N_s$ and $L_t = a N_t$, $a$ being the
lattice spacing, with $N_t \ll N_s$.  Periodic boundary conditions are
imposed in all directions. The temperature of the system is
then given by $T = 1/L_t$ and for fixed $N_t$ it becomes a
function of $\beta$ only, through the dependence of the lattice spacing $a$
on the gauge coupling. To find the value of the coupling at which the deconfining
transition takes place, at fixed $N_c$, we compute the
spatial average of the temporal Polyakov loop 
\begin{equation}
\bar{l}_p = \frac{1}{N_c N_s^3}\sum_{\vec{x}} \Tr
\left( \prod_{t = 0}^{N_t -1}  U_{4}(\vec{x},t)\right) \ ,
\end{equation}
where $\vec{x}$ and $t$ are the components of the Euclidean
four-vector $i$ respectively in the spatial and in the temporal
directions (the latter corresponding to the dimension of size $L_t$).
The deconfinement phase transition can be seen as a transition from the
phase symmetric under the center symmetry $\mathbb{Z}_N$ (the confined
phase), to the phase in which this symmetry is spontaneously
broken. $\bar{l}_p$ is the order parameter of the deconfinement phase
transition. In addition, we study
the four-volume average of the plaquette $\bar{u}_p$, defined as
\begin{equation}
  \label{eq:plaquette-lattice}
  \bar{u}_p = \frac{1}{N_c N_t N_s^3} \sum_{i,\mu > \nu} \mbox{Re} \ \Tr
    \left( U_{\mu \nu}(i) \right)
  \ .
\end{equation}
At fixed volume, we define the coupling $\beta_c(N_s,N_t)$ corresponding to the deconfinement
temperature by looking at the peak of the susceptibility of the
modulus of $\bar{l}_p$:
\begin{equation}
  \label{eq:sus-poly-loop}
  \chi_l \; = \; N_s^3 \left( \langle |\bar{l}_p|^2 \rangle - \langle
    |\bar{l}_p| \rangle^2 \right) \ .
\end{equation}
Our calculation is meant to complement the results already present in
the literature~\cite{Lucini:2002ku,Lucini:2003zr,Lucini:2005vg,Datta:2009jn}. Since
calculations at $N_c$ larger than $8$ become quite
expensive~\cite{Datta:2009jn}, we focused our attention to lower
$N_c$. For $N_c < 4$, very precise results are already available. For
$N_c = 4$ and $N_c = 6$, previous studies already attained a good level of
precision at $N_t=5,6,8$. We hence studied the case $N_t = 7$, which has not been
investigated before. The addition of these calculations helps to
improve the extrapolation of the corresponding $\beta_c$ to the continuum limit ($N_t =
\infty$). Calculations at $N_c = 5,7$ have not been performed
before. Hence, most of our numerical effort is devoted to the determination
of the critical temperature $T_c$ for SU(5) and SU(7). Finally, for
completeness, we have performed a new, high statistics numerical
investigation of SU(8) (comparable to that of~\cite{Datta:2009jn}),
which has enabled us to perform a more robust large $N_c$ extrapolation. 

\begin{table}[ht]
  \centering
  \begin{tabular}[h]{|c|c|c|c|}
    \hline
    \hline
    $N_c$ & $N_t$ & $N_s$ & $N_{\rm meas} \times 10^5$ \\
    \hline
    4 & 7 & 22 & 12 \\
    \hline
    5 & 5 & 8,10,12,14,16 & 3 \\
     & 6 & 14            & 2 \\
     & 7 & 16            & 6 \\
     & 8 & 18            & 5 \\
    \hline
    6 & 7 & 20 & 10 \\
    \hline
    7 & 5 & 8,10,11,12  & 2 \\
     & 6 & 10,12       & 3 \\
     & 7 & 11,12,13,14 & 6 \\
     & 8 & 14          & 6 \\
    \hline
    8 & 5 & 7,8,10,11 & 12 \\
     & 6 & 10        & 6  \\
     & 7 & 12        & 10 \\
     & 8 & 14        & 10 \\
    \hline \hline
  \end{tabular}
  \caption{Simulated volumes $N_s$ for different gauge groups
    SU($N_c$) and different temporal lengths $N_t$. An approximate
    counting of the total number of measurements on each lattice is
    also shown. The results on
    these new lattices complement and improve the study of Ref.~\cite{Lucini:2005vg}.}
  \label{tab:simulations-summary}
\end{table}
SU($N_c$) gauge theories for $N_c \ge 5$ have a bulk phase transition
at some value $\beta_B$ of the coupling constant. The continuum
physics is realised for $\beta > \beta_B$. For SU(5), $\beta_B \simeq
16.655$~\cite{Lucini:2005vg}. We have determined $\beta_B$ for SU(7), which
turns out to be around 33.246. The request that the system at
criticality be in the continuum regime is fulfilled if $N_t \ge
5$. As expected, this is the same bound on $N_t$ already found for
$N_c=4,6,8$. At fixed $N_c$, $N_s$ and $N_t$ we have computed  $\chi_l$ (see
Eq.~(\ref{eq:sus-poly-loop})) for about 10 $\beta$s in the critical
region, in a range that covers both the 
confined and the deconfined phase. For each calculation, we have used
a combination of overrelaxation and heath-bath updates with ratio
4:1. The number of sweeps has been chosen in such a way that at least
eight tunnellings were observed. In fact, in most of the cases we
observed 12--15 tunnellings for the largest lattices. 
The statistics for each
gauge group, spatial and temporal sizes is reported in
Tab.~\ref{tab:simulations-summary}. 
\begin{figure}[ht]
  \centering
  \includegraphics[width=0.45\textwidth]{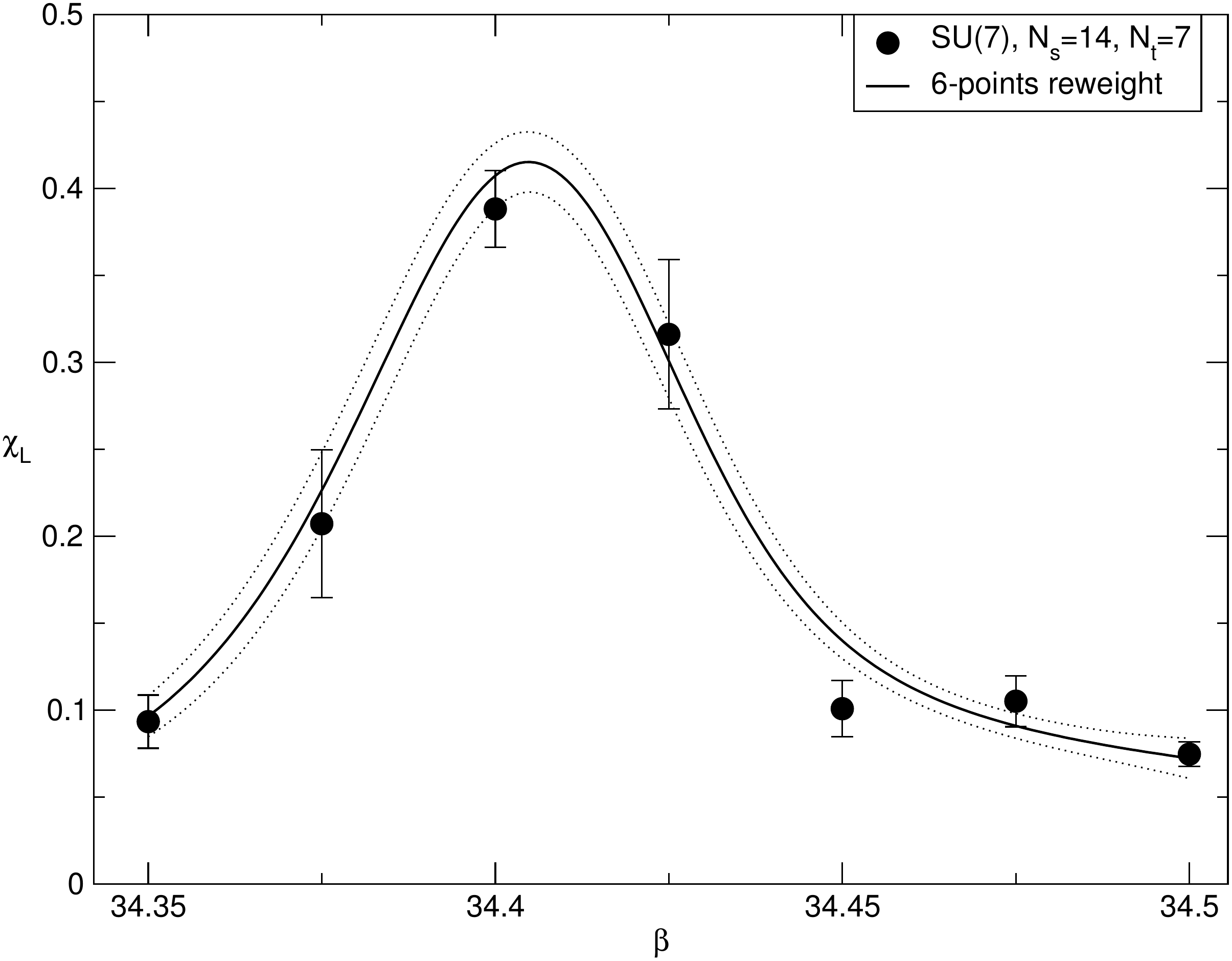}
  \caption{Reweighted data of $\chi_l$ for SU(7) gauge theory on  a
    $14^3 \times 7$ lattice. The filled circles are the measured data
    (with errors) and the unbroken line is the reweighted curve. The
    errors of the reweighted points (dashed lines) have been
    determined with a bootstrap (see details in the text).}  
  \label{fig:rew_sus_su7_l14_t7}
\end{figure}

Always at fixed $N_c$, $N_s$ and $N_t$, using the points that are
closer to the critical $\beta$ (typically five or six), we have
reweighted $\chi_l$ using the Ferrenberg-Swendsen procedure~\cite{Ferrenberg:1988yz}, as
illustrated for one set of parameters in
Fig.~\ref{fig:rew_sus_su7_l14_t7}.
In particular, the reweighted data have been generated for several
different bootstrap samples of the original simulations in order to give an
unbiased estimate of the statistical error. One $\beta$ value
corresponding to the location of the maximum is chosen for each bootstrap
sample (which is sufficiently well-behaved for a unique choice to be
made). The central value and the standard deviation of the Gaussian
distribution of these $\beta$ values determine our best estimate for the
critical coupling $\beta_c$. The procedure described above allows us
to avoid choosing the range for a quadratic fit approximating $\chi_l$
and therefore gives a more reliable and robust result.
\section{Thermodynamic limit}
\label{sect:thermodynamic}
For fixed $N_c$ and fixed $N_t$, the critical coupling
$\beta_c(N_t)$ is the thermodynamic limit of $\beta_c(N_s,N_t)$. Since
for $N_c \ge 3$ the phase transition is first
order~\cite{Lucini:2002ku,Lucini:2003zr}, the extrapolation is 
performed according to the ansatz 
\begin{equation}
  \label{eq:infinite-volume}
  \beta_c(\infty,N_t) \; = \; \beta_c(N_s,N_t) + h(N_t) \frac{N_t^3}{N_s^3} \ ,
\end{equation}
where only the leading volume correction is taken into account and the
value of the coefficient $h(N_t)$ depends on the lattice 
spacing~\cite{Lucini:2005vg}:
\begin{equation}
h(N_t) \mathop{=}_{a \to 0} 
\frac{h_0}{K\left(\beta_c(N_t)\right)} + {\cal O}(a^2)  \ ,
\end{equation}
with
\begin{equation}
K\left(\beta_c(N_t)\right) = \left. \frac{\mathrm{d} \ln a
    (\beta)}{\mathrm{d} \beta}\right|_{\beta = \beta_c(N_s,N_t)} \ .
\end{equation}
The procedure to evaluate $h(N_t)$ can be described as follows.
At first we obtain the coefficient $h(N_t=5)$ (corresponding to our largest lattice spacing)
directly from a finite--size scaling (FSS) analysis using a wide range
of volumes $V=N_s^3$.
We then use our own data for the beta function
$\frac{\partial a(\beta)}{\partial \beta}$ obtained from the
interpolation of the string tension over a 
large set of couplings in order to estimate $h(N_t)$ at $N_t=6,7,8$, with higher 
order corrections ${\cal O}(a^2)$ accounted for by a $15\%$ error
increase on $h(N_t)$~\cite{Lucini:2005vg}.

The determination of the beta function requires measuring
zero-momentum correlators of Po\-lya\-kov loops at zero temperature
({\em torelons}). In order to extract $a m_l$, the
mass of the loop in lattice units, we look at the large time
separation exponential decay of zero-momentum correlators of spatial
Polyakov loops, which is controlled by $a m_l$ itself (see
e.g. Ref.~\cite{Lucini:2004my}).
For SU($N_c$) gauge groups with $N_c=2,3,4,6,8$,
detailed measurements of torelon masses $am_l$ are already available on a
wide range of coupling constants~\cite{Lucini:2004my}. From the mass
of such states extracted using spatial Po\-lya\-kov loops of length $N_s$ in units
of the lattice spacing, we obtain the string tension $a\sqrt{\sigma}$
by solving the equation
\begin{equation}
  \label{eq:string-corrections}
  am_l(N_s) \; = \; a^2\sigma N_s - \frac{\pi}{3N_s} -
  \frac{\pi^2}{18N_s^3}\frac{1}{a^2\sigma}
  \ ,
\end{equation}
where the last two terms immediately derive from the effective theory
describing the low--energy dynamics of confining strings in the
SU($N_c$) theory~\cite{Aharony:2009gg}. If we keep $\sigma$ fixed to
its continuum value, the numerical data for $a \sqrt{\sigma}$ give us
the variation of $a$ as a function of $\beta$.
\begin{table}[ht]
  \centering
  \begin{tabular}[h]{|c|c|c|c|}
    \hline
    \hline
    $N_c$ & $L$ & $\beta$ & $a\sqrt{\sigma}$ \\
    \hline
    4 & 14 & 10.9415 & 0.2314(11) \\
    \hline
    5 & 10 & 16.8762 & 0.3352(17)  \\   
     & 12 & 17.1070 & 0.2755(10)  \\  
     & 14 & 17.22   & 0.25530(74) \\  
     & 14 & 17.3371 & 0.23649(53) \\  
     & 16 & 17.44   & 0.22093(58) \\
     & 16 & 17.556  & 0.20710(53) \\  
     & 18 & 17.66   & 0.19386(40) \\
    \hline
    6 & 14 & 25.1707 & 0.2379(9) \\
    \hline
    7 & 10 & 33.5465 & 0.3439(22)  \\
     & 12 & 33.9995 & 0.27981(96) \\
     & 14 & 34.22   & 0.25950(75) \\
     & 14 & 34.4397 & 0.23997(76) \\
     & 16 & 34.63   & 0.22435(47) \\
     & 16 & 34.8295 & 0.21010(68) \\
     & 18 & 35.00   & 0.2011(10)  \\
    \hline
    8 & 14 & 44.0955 & 0.2426(6) \\
    \hline \hline
  \end{tabular}
  \caption{SU(5) and SU(7) string tensions on hypercubic
    lattices $L^4$ at the reported values of $\beta$. The string
    tension is extracted from the mass of the lightest torelon state
    of length $L$. SU(4), SU(6) and SU(8) string tensions at the
    critical coupling for $N_t=7$ are also shown.} 
  \label{tab:string-tension}
\end{table}

In Tab.~\ref{tab:string-tension} we summarise the string tension
measured in high statistics simulations on large symmetric lattices
$L^4$ at the reported values of $\beta$. For the computation of
$\sigma$, we have used Eq.~(\ref{eq:string-corrections}). Since
previous lattice calculations only used the leading correction  $-
\frac{\pi}{3N_s}$, we have verified that the insertion of the
next-to-leading correction, whose universal character has been
discovered only recently~\cite{Aharony:2009gg,Luscher:2004ib}, does not
affect the numerical results within errors.  We have calculated for the
first time the behaviour of the string tension in SU(5) and SU(7) as a
function of the bare coupling. This allows us to have a precise
estimate of the string tension at the couplings corresponding to the
deconfinement temperature for $N_t=5,6,7,8$. In addition to the above
gauge groups, we obtained the string tension for SU(4), SU(6) and
SU(8) in the neighbourhood of the critical coupling $\beta_c$ for
$N_t=7$, needed for the improvement of the continuum extrapolation of
the critical temperature.

For each value of $N_c \ge 4$, we interpolated the larger set of
string tensions available to us using a polynomial function to obtain
$a(\beta)\sqrt{\sigma}$. This interpolation includes a nested
bootstrap sampling for a better error estimation and agrees very
well with the fits of older data performed in
Ref.~\cite{Lucini:2005vg} (for an alternative procedure, see
also Ref.~\cite{Allton:2008ty}). Using $a(\beta)\sqrt{\sigma}$ and its
derivative with respect to the coupling we obtain the infinite volume
extrapolation $\beta_c(\infty,N_t)$ at $N_t=6,7,8$ shown in
Tab.~\ref{tab:beta-critical}. 
\begin{table}[ht]
  \centering
  \begin{tabular}[h]{|c|c|c|c|c|}
    \hline
    \hline
    $N_c$ & $N_t$ & $N_s$ & $h(N_t)$ & $\beta_c$ \\
    \hline
    4 & 7 & 22    & 0.111(24) & 10.9415(12) \\
    \hline
    5 & 5 & 12-16 & 0.129(23) & 16.8762(12) \\
     & 6 & 14 & 0.138(28) & 17.1074(33)    \\
     & 7 & 16 & 0.147(30) & 17.3386(31)    \\
     & 8 & 18 & 0.157(32) & 17.5585(36)    \\
    \hline
    6 & 7 & 20    & 0.149(29) & 25.1715(26) \\
    \hline
    7 & 5 & 10-12 & 0.114(17) & 33.5465(11) \\
     & 6 & 12 & 0.162(14) & 34.0001(38)    \\
     & 7 & 14 & 0.167(15) & 34.4256(29)    \\
     & 8 & 14 & 0.171(15) & 34.8318(84)    \\
    \hline
    8 & 5 & 8-11 & 0.150(10) & 43.9793(16) \\
     & 6 & 10  & 0.175(13)  & 44.5556(65) \\
     & 7 & 12  & 0.207(16)  & 45.1145(42) \\
     & 8 & 14  & 0.248(19)  & 45.6438(49) \\
    \hline \hline
  \end{tabular}
  \caption{SU(5), SU(7) and SU(8) values of the critical
    coupling for the corresponding temporal extent and in the infinite
    volume limit. Simulations are performed on $N_s^3\times N_t$ lattices and
    the values $\beta_c(\infty,N_t)$ are obtained from
    Eq.~(\ref{eq:infinite-volume}). For SU(4) and SU(6) we also report
    our novel estimate of $\beta_c$ at fixed $N_t=7$.}
  \label{tab:beta-critical}
\end{table}

As a final remark on the thermodynamic limit, we comment on the possible influence of the bulk phase transition. As it has been pointed out in Ref.~\cite{Datta:2009jn}, at
the lattice spacing corresponding to $\beta_c$ for $N_t=5$ the theory
might still be affected by artefacts related to the nearby bulk phase
transition. In the SU(7) gauge theory at $N_t = 5$, for which the bulk phase transition is very close to the finite-temperature transition, we have explicitly checked whether the
determination of $h(N_t)$ at that lattice spacing is significantly affected
by lattice artefacts. To this purpose, we performed a FSS analysis on
lattices with a larger temporal extension ($N_t=7$), where the deconfinement transition is pushed to
a weaker coupling. Both procedures give infinite volume
estimates for $\beta_c$ that are compatible within the statistical
uncertainty of our simulations.
\section{Continuum extrapolation}
\label{sect:continuum}
Before we can take the continuum and the large $N_c$ limit, 
the location of the deconfinement transition we found in the previous
section needs to be translated into a physical temperature $T_c$. In addition, 
for the continuum limit, it proves convenient to use dimensionless
quantities, since at the leading order these have corrections that are
quadratic in the lattice spacing. Using the string tension $\sqrt{\sigma}$
to set the scale of pure gauge lattice simulations gives good control
over systematic errors in the continuum extrapolation.  For this reason, we study the continuum limit of the
deconfinement temperature in units of the square root of $\sigma$, 
$T_c/\sqrt{\sigma}$. This is determined for each $N_c$ using only the leading
${\cal O}(a^2)$ correction~\cite{Lucini:2003zr}:
\begin{equation}
  \label{eq:continuum-limit-tcsigma}
  \left. \frac{T_c}{\sqrt{\sigma}} \right|_{a = 0}\; = \;
  \left. \frac{T_c}{\sqrt{\sigma}} \right|_{a} + \delta a^2\sigma \ ,
\end{equation}
where at fixed lattice spacing
\begin{equation}
  \label{eq:fixed-a-tcsigma}
  \left. \frac{T_c}{\sqrt{\sigma}}\right|_a \; = \;
  \frac{1}{N_ta(\beta_c)\sqrt{\sigma}} 
\end{equation}
and $\delta$ is a numerical coefficient of order one.
We performed the continuum limit of $T_c/\sqrt{\sigma}$ according to
Eq.~(\ref{eq:continuum-limit-tcsigma}) and using four different
lattice spacings. The precision we achieve on the ratio $T_c/\sqrt{\sigma}$ is
mostly determined by the precision of the string tension, since this latter
quantity is affected by a relative error larger than that of $\beta_c$.  

\begin{figure}[ht]
  \centering
  \includegraphics[width=0.45\textwidth]{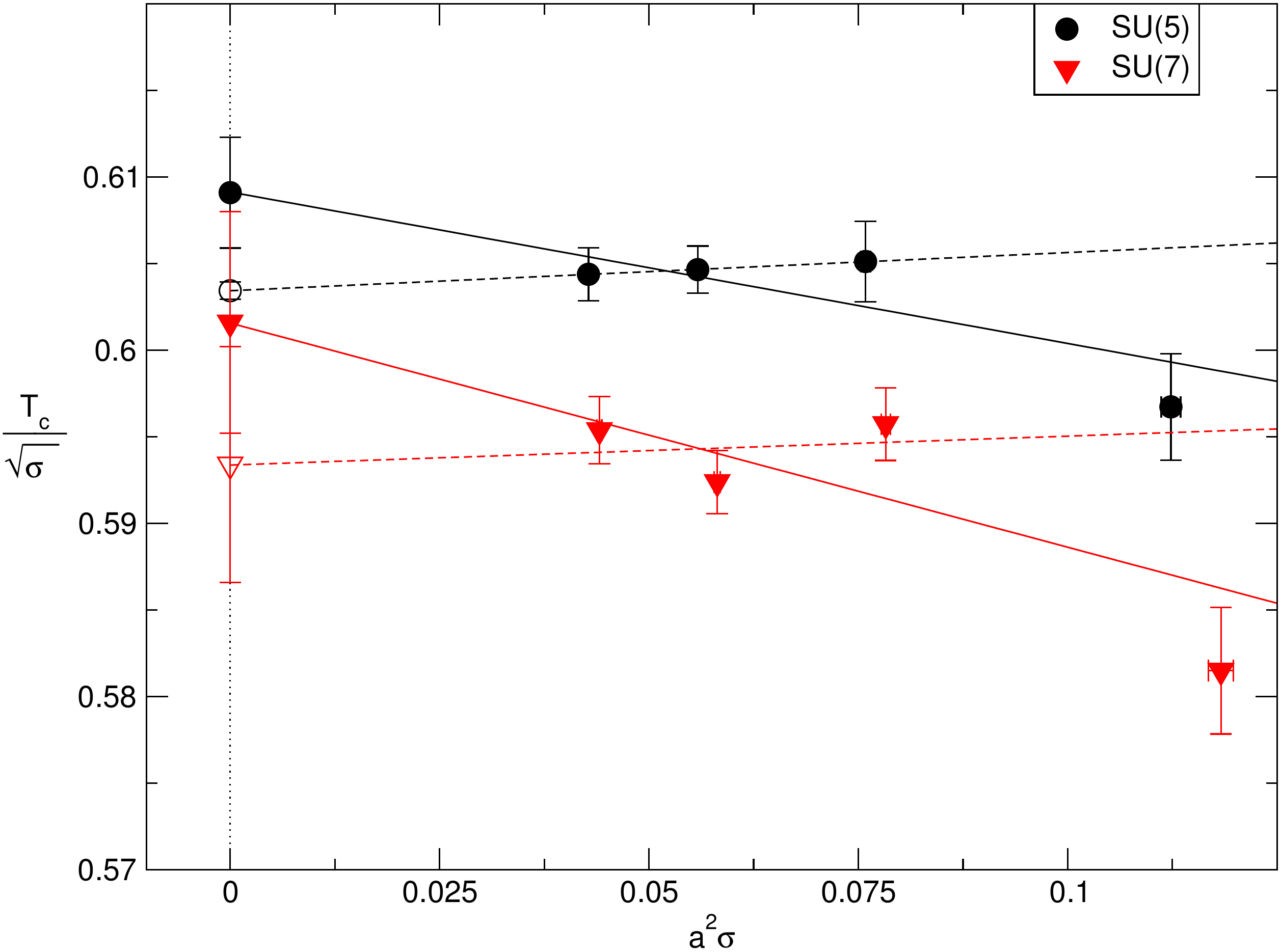}
  \caption{Continuum extrapolation of $T_c/\sqrt{\sigma}$ for SU(5)
    and SU(7) according to Eq.~(\ref{eq:continuum-limit-tcsigma}) and
    with different fitting ranges. The continuum values are shown on
    the left. The dashed line, together with the open symbols,
    correspond to extrapolations without the coarsest lattice point.}
  \label{fig:continuum-su5-su7}
\end{figure}

The availability of an additional lattice
spacing in the asymptotic scaling region for the continuum
extrapolation can help us identifying possible systematic effects due
to the inclusion of the coarsest 
point. In Fig.~\ref{fig:continuum-su5-su7} we show the continuum
limit of the deconfinement temperature for SU(5) and SU(7), which is the key original
contribution of this work. Fits with and without the coarsest lattice
point give compatible results for SU(7) with a $\chi^2$ per
d.o.f. above three disfavouring the former. For SU(5) the situation
is similar, but the fit with all points has an acceptable $\chi^2$.
This is also true for $N_c = 4, 6, 8$. Our conclusion is that if there
is any systematic effect in extrapolating the
ratio $T_c/\sqrt{\sigma}$ to the $a = 0$ limit including points measured on $N_t = 5$
lattices, this effect is significantly smaller than the statistical
error.

In Tab.~\ref{tab:sun-tcsigma-continuum} we summarise the continuum 
limit values of $T_c/\sqrt{\sigma}$ that we used to obtain the
SU($\infty$) limit. Only for $N_c=7$ we discard the coarsest lattice
point in the continuum limit and show the result for the fit obtained
using $N_t > 5$. Since all the other gauge groups have well-behaved
extrapolations with a low $\chi^2$ even including the $N_t = 
5$ point, in those cases we perform the fit using results at all the
available values of $N_t$.

\section{Large $N_c$ extrapolation}
\label{sect:tclargen}
\begin{figure}[ht]
  \centering
  \includegraphics[width=0.45\textwidth]{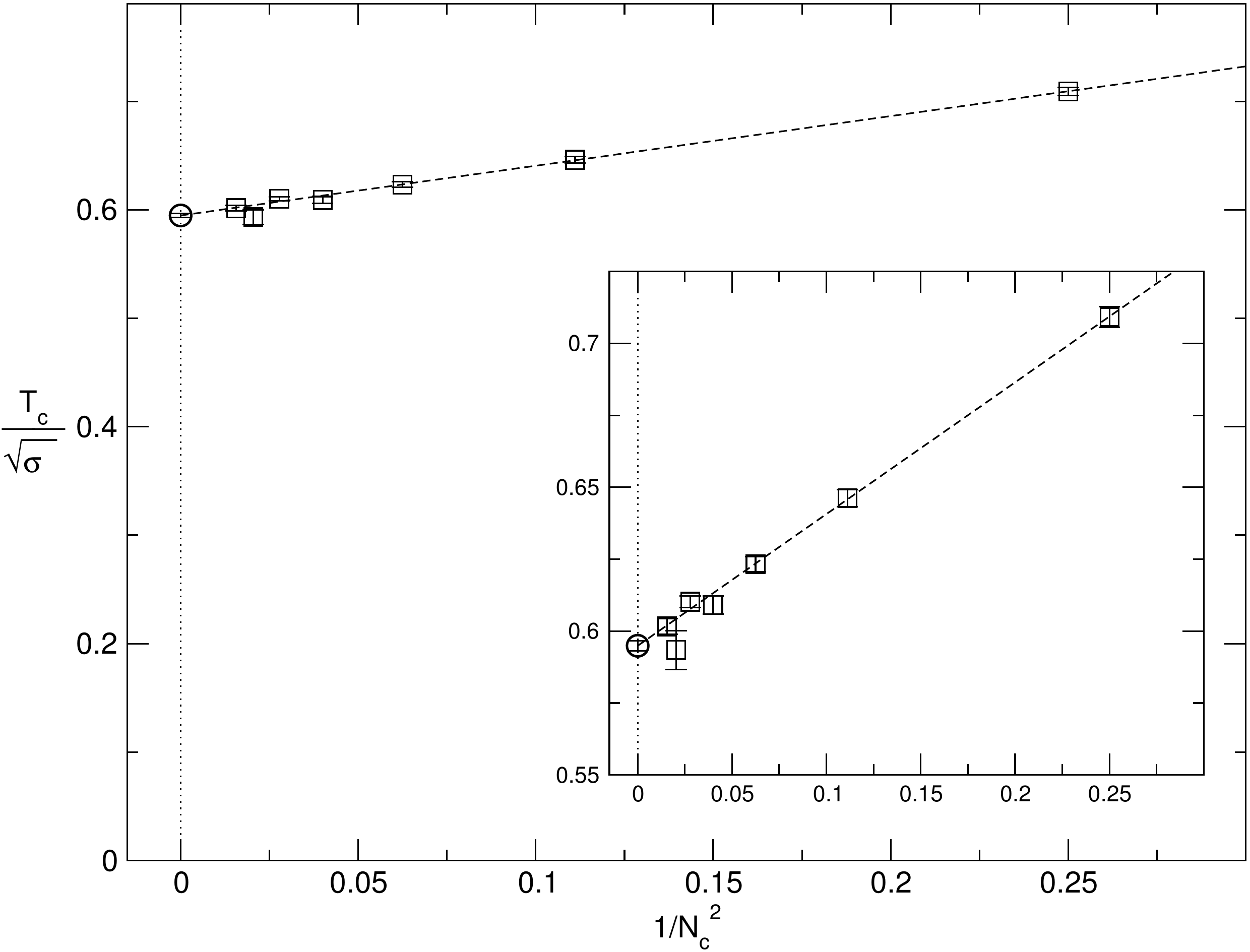}
 \caption{Large $N_c$ extrapolation of $T_c/\sqrt{\sigma}$ using all
    data in Tab.~\ref{tab:sun-tcsigma-continuum}. The dashed line
    corresponds to the fitted formula in
    Eq.~(\ref{eq:largeN-limit}). The inset is a close-up on the data.}
  \label{fig:largeN}
\end{figure}

According to large $N_c$ arguments, the large $N_c$ limit of
$T_c/\sqrt{\sigma}$ can be expressed as a power series in $1/N_c^2$. 
\begin{eqnarray}
\label{eq:largenc-ansatz}
\left. \frac{T_c}{\sqrt{\sigma}} \right|_{N_c}= 
\left. \frac{T_c}{\sqrt{\sigma}} \right|_{N_c = \infty} +
\frac{c}{N_c^2} + {\cal O}(N_c)^{-4} \ ,
\end{eqnarray}
where $c$ is a numerical constant of order one. Our best fit for the large
$N_c$ deconfinement temperature according to
Eq.~(\ref{eq:largenc-ansatz}) is shown in Fig.~\ref{fig:largeN}. We
extrapolate keeping only the $1/N_c^2$ correction to the planar limit,
a procedure that has been shown to work very well down to $N_c=2$~\cite{Lucini:2005vg}.
Including also SU(2) and SU(3) data from Ref.~\cite{Lucini:2003zr}, we obtain
\begin{equation}
  \label{eq:largeN-limit}
  T_c/\sqrt{\sigma} = 0.5949(17) + 0.458(18)/N_c^2
  \ ,
\end{equation}
with good $\chi^2/{\rm d.o.f.} = 1.18$. Discarding $N_c=2,3$ worsens the
quality of the fit without changing the fitted parameters
within the quoted error. Our value for the SU($\infty$) deconfinement
temperature in units of the string tension is compatible with previous
results reported in Ref.~\cite{Lucini:2005vg}, but the relative accuracy has
increased approximately by a  factor of $2$. This is due to a better control over
the continuum extrapolation for $N_c \ge 4$ and to the inclusion of
SU(5) and SU(7) data. Note that the more precise result is still
compatible with the finite $N_c$ value being accounted for by the leading
$1/N_c^2$ correction only.

\begin{table}[t]
  \centering
  \begin{tabular}[h]{|c|c|c|}
    \hline \hline
    \multicolumn{3}{|c|}{SU($N_c$)}\\
    \hline
    $N_c$ & $T_c/\sqrt{\sigma}$ & $\chi^2/{\rm d.o.f}$ \\
    \hline
    2 & 0.7092(36) & 0.28 \\
    3 & 0.6462(30) & 0.05 \\
    4 & 0.6233(26) & 0.69 \\
    5 & 0.6091(32) & 1.25 \\
    6 & 0.6102(20) & 0.26 \\
    7 & 0.5934(68) & 1.8  \\
    8 & 0.6016(27) & 0.69 \\
    \hline \hline
    $\infty$ & 0.5949(17) & 1.18 \\
    \hline \hline
  \end{tabular}
  \caption{Critical temperature in units of the string tension in the
    continuum limit for different gauge groups. The SU(2) and
    SU(3) values are taken from Ref.~\cite{Lucini:2005vg}. The
    large $N_c$ extrapolation using all the reported values is shown
    in the bottom row.}
  \label{tab:sun-tcsigma-continuum}
\end{table}
In order to assess the robustness of our result, we tested it against
possible systematic errors in the continuum extrapolation. A different
set of continuum values was created in the
following way: for each $N_c \ge 4$, results of continuum fits with and
without the $N_t=5$ point were merged together such that the error
accounted for the whole possible range of values, while the middle
point of the error bar was taken as the central value. The estimates we obtained
are fully compatible with Eq.~(\ref{eq:largeN-limit}). Fitting
the large $N_c$ behaviour of points obtained by extrapolating to the
continuum limit results for $N_t \ge 6$ for all $N_c \ge 4$ (except
for $N_c = 5$, where if we consider only points at $N_t > 5$ the
errors resulting from the continuum fit are anomalously small) also gives
compatible results. 
\section{Conclusions}
\label{sect:conclusions}
We have determined numerical values of the ratio $T_c/\sqrt{\sigma}$ for gauge groups
SU(5) and SU(7). We have used the new data together with results
for $N_c = 2, 3, 4, 6, 8$ already available in the
literature~\cite{Lucini:2002ku,Lucini:2003zr,Lucini:2005vg}
(supplemented with calculations at an additional lattice spacing for
$N_c = 4,6$ and with calculations with increased statistics for $N_c = 8$) to
reanalyse the large $N_c$ limit of this quantity, for which it turns
out that only the leading $1/N_c^2$ correction is needed to
extrapolate finite $N_c$ results in the range $2 \le N_c \le 8$. This
had been already observed in previous simulations. We obtain an
accurate large $N_c$ limit that 
improves by a factor of two the precision of previous calculations. At
the same time, we investigated possible finite lattice spacing artefacts. Our
analysis lead us to the conclusion that for $4 \le N_c \le 8$, it is safe to extrapolate to
the continuum limit from $N_t = 5$, as done
in~\cite{Lucini:2002ku,Lucini:2003zr}. 

In order to obtain a further noticeable improvement, it is likely that
gauge groups with $N_c \ge 8$ need to be investigated. However, since
the strength of the first order deconfinement transition grows with
$N_c$, reliable Monte Carlo studies of those systems will crucially require
algorithms that mitigates substantially the exponential suppression in the
spatial volume of the tunnelling rate between the confined and the
deconfined phases at criticality, like for instance the multicanonical
algorithm~\cite{Berg:1992qua}. 

\section*{Acknowledgments}
B.L. acknowledges financial support from the Royal Society (grant UF09003)
and STFC (grant ST/G000506/1). The simulations discussed in this Letter have
been performed on a cluster partially funded by STFC and by the Royal Society.
E.R. is funded by a SUPA Prize Studentship. E.R. acknowledges hospitality
and financial support from the INFN, Laboratori Nazionali di Frascati,
during the final stage of this work.





\bibliographystyle{h-elsevier3.bst}
\bibliography{suntc.bib}

\begin{thebibliography}{10}

\bibitem{Teper:2009uf}
M. Teper,
\newblock Acta Phys.Polon. B40 (2009) 3249, 0912.3339.

\bibitem{Narayanan:2009xh}
R. Narayanan,
\newblock Acta Phys.Polon. B40 (2009) 3231, 0910.3711.

\bibitem{Hooft:1973jz}
G. 't~Hooft,
\newblock Nucl. Phys. B72 (1974) 461.

\bibitem{Witten:1979kh}
E. Witten,
\newblock Nucl. Phys. B160 (1979) 57.

\bibitem{Manohar:1998xv}
A.V. Manohar,
\newblock (1998), hep-ph/9802419.

\bibitem{Aharony:1999ti}
O. Aharony et~al.,
\newblock Phys. Rept. 323 (2000) 183, hep-th/9905111.

\bibitem{Lucini:2002ku}
B. Lucini, M. Teper and U. Wenger,
\newblock Phys. Lett. B545 (2002) 197, hep-lat/0206029.

\bibitem{Lucini:2003zr}
B. Lucini, M. Teper and U. Wenger,
\newblock JHEP 01 (2004) 061, hep-lat/0307017.

\bibitem{Lucini:2004yh}
B. Lucini, M. Teper and U. Wenger,
\newblock Nucl. Phys. B715 (2005) 461, hep-lat/0401028.

\bibitem{DelDebbio:2004rw}
L. Del~Debbio, H. Panagopoulos and E. Vicari,
\newblock JHEP 09 (2004) 028, hep-th/0407068.

\bibitem{deForcrand:2004jt}
P. de~Forcrand, B. Lucini and M. Vettorazzo,
\newblock Nucl.Phys.Proc.Suppl. 140 (2005) 647, hep-lat/0409148.

\bibitem{Bursa:2005yv}
F. Bursa and M. Teper,
\newblock JHEP 0508 (2005) 060, hep-lat/0505025.

\bibitem{Lucini:2005vg}
B. Lucini, M. Teper and U. Wenger,
\newblock JHEP 02 (2005) 033, hep-lat/0502003.

\bibitem{Bringoltz:2005rr}
B. Bringoltz and M. Teper,
\newblock Phys. Lett. B628 (2005) 113, hep-lat/0506034.

\bibitem{Bringoltz:2005xx}
B. Bringoltz and M. Teper,
\newblock Phys.Rev. D73 (2006) 014517, hep-lat/0508021.

\bibitem{deForcrand:2005rg}
P. de~Forcrand, B. Lucini and D. Noth,
\newblock PoS LAT2005 (2006) 323, hep-lat/0510081.

\bibitem{Liddle:2005qb}
J. Liddle and M. Teper,
\newblock PoS LAT2005 (2006) 188, hep-lat/0509082.

\bibitem{Liddle:2008kk}
J. Liddle and M. Teper,
\newblock (2008), 0803.2128.

\bibitem{Kiskis:2009xj}
J. Kiskis and R. Narayanan,
\newblock Phys.Lett. B679 (2009) 535, 0906.3015.

\bibitem{Panero:2009tv}
M. Panero,
\newblock Phys. Rev. Lett. 103 (2009) 232001, 0907.3719.

\bibitem{Datta:2009jn}
S. Datta and S. Gupta,
\newblock Phys. Rev. D80 (2009) 114504, 0909.5591.

\bibitem{Caselle:2011fy}
M. Caselle et~al.,
\newblock JHEP 1106 (2011) 142, 1105.0359.

\bibitem{Caselle:2011mn}
M. Caselle et~al.,
\newblock (2011), 1111.0580.

\bibitem{Mykkanen:2012ri}
A. Mykkanen, M. Panero and K. Rummukainen,
\newblock (2012), 1202.2762.

\bibitem{Lucini:2004my}
B. Lucini, M. Teper and U. Wenger,
\newblock JHEP 06 (2004) 012, hep-lat/0404008.

\bibitem{DelDebbio:2007wk}
L. Del~Debbio et~al.,
\newblock JHEP 03 (2008) 062, 0712.3036.

\bibitem{Armoni:2008nq}
A. Armoni et~al.,
\newblock Phys.Rev. D78 (2008) 045019, 0804.4501.

\bibitem{Lucini:2010nv}
B. Lucini, A. Rago and E. Rinaldi,
\newblock JHEP 1008 (2010) 119, 1007.3879.

\bibitem{Lucini:2010kj}
B. Lucini et~al.,
\newblock Phys.Rev. D82 (2010) 114510, 1008.5180.

\bibitem{Ferrenberg:1988yz}
A. Ferrenberg and R. Swendsen,
\newblock Phys.Rev.Lett. 61 (1988) 2635.

\bibitem{Aharony:2009gg}
O. Aharony and E. Karzbrun,
\newblock JHEP 06 (2009) 012, 0903.1927.

\bibitem{Luscher:2004ib}
M. Luscher and P. Weisz,
\newblock JHEP 07 (2004) 014, hep-th/0406205.

\bibitem{Allton:2008ty}
C. Allton, M. Teper and A. Trivini,
\newblock JHEP 07 (2008) 021, 0803.1092.

\bibitem{Berg:1992qua}
B. Berg and T. Neuhaus,
\newblock Phys.Rev.Lett. 68 (1992) 9, hep-lat/9202004.

\end{thebibliography}







\end{document}